\documentclass[aps,pra,twocolumn]{revtex4}%
\usepackage{graphics}
\usepackage{amsmath}
\usepackage{graphicx}
\usepackage{amsfonts}
\usepackage{amssymb}%
\begin{document}
\title{CV-MDI-QKD with coherent state: One-mode Gaussian attacks are not enough}
\author{Carlo Ottaviani}
\affiliation{Department of Computer Science and York Centre for Quantum Technologies,
University of York, York YO10 5GH, United Kingdom}
\author{Gaetana Spedalieri}
\affiliation{Department of Computer Science and York Centre for Quantum Technologies,
University of York, York YO10 5GH, United Kingdom}
\author{Samuel L. Braunstein}
\affiliation{Department of Computer Science and York Centre for Quantum Technologies,
University of York, York YO10 5GH, United Kingdom}
\author{Stefano Pirandola}
\affiliation{Department of Computer Science and York Centre for Quantum Technologies,
University of York, York YO10 5GH, United Kingdom}

\begin{abstract}
The security proof of continuous variable (CV) measurement device independent
(MDI) quantum key distribution (QKD) cannot be reduced to the analysis of
one-mode Gaussian attacks (in particular, independent entangling-cloner
attacks). To stress this point, the present work provides a very simple
(almost trivial) argument, showing that there are an infinite number of
two-mode Gaussian attacks which cannot be reduced to or simulated by one-mode
Gaussian attacks. This result further confirms that the security analysis of
CV-MDI-QKD must involve a careful minimization over two-mode attacks as
originally performed in [S. Pirandola \textit{et al.}, Nature Photon.
\textbf{9}, 397-402 (2015)].

\end{abstract}
\maketitle

\section{Introduction}

Measurement-device-independent quantum key distribution MDI-QKD
\cite{SidePRL,Lo} promises to be a remarkably effective solution for the
practical implementation of the next generation of QKD infrastructures, in
which privacy should be granted over a quantum network. In MDI-QKD the
authorized users of the network, Alice and Bob, exploit a swapping-like
protocol where secret correlations are established by the measurement of a
third untrusted party, the relay \cite{SidePRL,CVMDIQKD,Carlo}. This performs
a Bell measurement but, in order to achieve security, it is not required to
pass a Bell test. By contrast, in full device independent QKD
\cite{EKERT,ACIN-DI}, the privacy of the shared key depends on passing a Bell
test, which is still an operation performed with very poor success rates
\cite{VAZIRANI,COLBECK,BELL-TEST}. The power of the MDI approach relies indeed
on its practicality: One can achieve high-rate side-channel-free
unconditionally secure network communication.

In recent years the study of QKD protocols based on quantum
continuous variables (CVs) \cite{diamanti-leverrier} has attracted
increasingly attention because of several appealing properties of
CV systems: Protocols use bright coherent states, and exploit
standard telecommunication technologies; in particular coherent
detection techniques, already developed for classical optical
communication \cite{RALPH}. In addition, CV-QKD is interesting for
the relatively simple implementation of protocols at different
frequencies \cite{weed2way,FILIP}. Finally, exploiting CV
point-to-point protocol with state-of-the-art classical
reconciliation and error correction schemes
\cite{leverrier,leverrier2} allowed the \textit{in field}
implementation CV-QKD over a distance of $80$ Km
\cite{jouguet-NAT-PHOT}.

In recent works \cite{CVMDIQKD,Carlo,ENT-REACT} we proposed a
CV-MDI-QKD protocol, which we have also successfully tested in a
proof-of-principle experiment \cite{CVMDIQKD}. In particular, we
proved that our scheme is capable of remarkably high key-rates per
use of the communication channel, over the length of metropolitan
range distances. This performance is orders of magnitude higher
than comparably practical implementations based on discrete
variable \cite{reply-NAT-PHOT}. We therefore believe that CV-QKD
will play a crucial role in future implementation of metropolitan
quantum cryptography. At this scale, in fact, both high density of
untrusted nodes and high rates should be considered nonnegotiable
properties, if we want a quantum network able of competing with
present classical infrastructure.

In this work we provide additional evidences supporting the security analysis
of CV-MDI-QKD given in Refs.~\cite{CVMDIQKD,Carlo}. We show, by simple
arguments, that the security analysis restricted to one-mode Gaussian
(entangling-cloner) attacks can only account for a subclass of all possible
eavesdroppings. In particular we provide a counterexample in order to
explicitly prove that, if we model Eve's attack assuming a restricted
strategy, based on independent entangling cloners, one cannot generate all the
possible covariance matrices shared between Alice and Bob. Our analysis
confirms that a complete security analysis of CV-MDI-QKD cannot indeed avoid
to consider two-mode Gaussian attacks, as originally done in
Refs.~\cite{CVMDIQKD,Carlo}.

The structure of the paper is the following. In
Section~\ref{protocol-DESCRIPTION} we present the protocol.
Section~\ref{GENERAL-CONSIDERATION} gives general consideration about the
security analysis, marking the difference between theoretical and experimental
analyses. Section~\ref{COUNTER-EXAMPLE}\ provides a simple counter example to
the (wrong) assumption that an attack by independent entangling cloners would
be complete. Finally, Sec.~\ref{CONCLUSIONS} is for our conclusions.

\section{Description of the protocol\label{protocol-DESCRIPTION}}

\begin{figure}[ptbh]
\vspace{-1.8cm}
\par
\begin{center}
\includegraphics[width=0.5\textwidth] {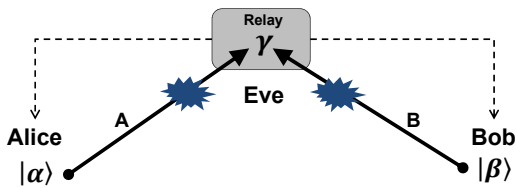}
\end{center}
\par
\vspace{-2.3cm}\caption{Basic protocol of CV-MDI-QKD}%
\label{PMscheme}%
\end{figure}

We start with a brief description of the protocol~\cite{CVMDIQKD}. At one
side, Alice prepares a mode $A$ in a coherent state $\left\vert \alpha
\right\rangle $ whose amplitude $\alpha$ is modulated by a Gaussian
distribution with zero mean and large variance. At the other side, Bob
prepares his mode $B$ in another coherent state $\left\vert \beta\right\rangle
$ whose amplitude $\beta$ is modulated by the same Gaussian distribution as
Alice. Modes $A$ and $B$ are then sent to an intermediate relay where a CV
Bell detection is performed. The classical outcomes are combined in a complex
variable $\gamma$, which is communicated to Alice and Bob via a public
channel. As a result, knowledge of $\gamma$ enables each party to infer the
variable of the other party by simple postprocessing (see Fig.~\ref{PMscheme}%
). 

In general, the relay is assumed to be untrusted~\cite{SidePRL}, i.e.,
operated by Eve, and also the links with the relay are subject to
eavesdropping. The protocol is assumed to be performed many times, so that the
honest parties collect a large amount of classical data (we consider
asymptotic security here). Using several tools, including de Finetti arguments
and the extremality of Gaussian states (see Ref.~\cite{CVMDIQKD}\ for more
details), one can reduce the security analysis to considering a two-mode
Gaussian attack against the two links with the relay (performing a proper CV
Bell detection). This type of attack can be constructed by suitably combining
two canonical forms~\cite{RMP} into a correlated-noise Gaussian environment.
The most relevant canonical forms are clearly the lossy channels.

\begin{figure}[ptbh]
\vspace{-1.99cm}
\par
\begin{center}
\includegraphics[width=0.5\textwidth] {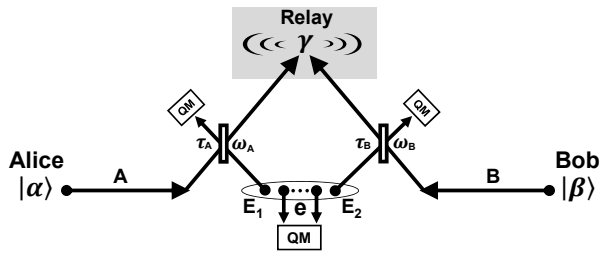}
\end{center}
\par
\vspace{-2.1cm}\caption{Two-mode Gaussian attack against CV-MDI-QKD. Figure
adapted from Ref.~\cite{CVMDIQKD}.}%
\label{PMscheme2}%
\end{figure}

In this scenario, the two modes $A$ and $B$ are mixed with two ancillary
modes, $E_{1}$ and $E_{2}$, by two beam splitters with transmissivities
$\tau_{A}$ and $\tau_{B}$, respectively. These ancillary modes belong to a
reservoir of ancillas ($E_{1}$, $E_{2}$ plus an extra set $\mathbf{e}$) in a
pure Gaussian state. The reduced state $\sigma_{E_{1}E_{2}}$ is a correlated
thermal state with zero mean and covariance matrix (CM) in the normal form%
\begin{equation}
\boldsymbol{\sigma}_{E_{1}E_{2}}=\left(
\begin{array}
[c]{cc}%
\omega_{A}\mathbf{I} & \mathbf{G}\\
\mathbf{G} & \omega_{B}\mathbf{I}%
\end{array}
\right)  ,~\mathbf{G}:=\left(
\begin{array}
[c]{cc}%
g & 0\\
0 & g^{\prime}%
\end{array}
\right)  , \label{EveCM_E1E2}%
\end{equation}
where $\omega_{A},\omega_{B}\geq1$ are the variances of the thermal noise
affecting each link, while $g$ and $g^{\prime}$ are correlation parameters,
satisfying suitable physical constraints~\cite{NJP2013,TwomodePRA}. After
interaction, Eve's ancillas are stored in a quantum memory, measured at the
end of the protocol (see Fig.~\ref{PMscheme2}). 

In order to deal with the joint attack, Alice and Bob must retrieve the joint
statistics of the variables $\alpha$, $\beta$, and $\gamma$. For this purpose,
they publicly compare a small part of their data and reconstruct the
probability distribution $p(\alpha,\beta,\gamma)$. The empirical values of the
transmissivities $\tau_{A}$ and $\tau_{B}$ are accessible to the parties from
the first-order moments of $p(\alpha,\beta,\gamma)$. Knowing these values is
essential in order to apply the correct post-processing and re-scaling of the
output data. Then, from the second-order moments of $p(\alpha,\beta,\gamma)$,
Alice and Bob can extract the CM $\mathbf{V}_{ab|\gamma}$ that they would
share in an equivalent entanglement-based representation of the
protocol~\cite{EBrepr} and conditioned to the outcome $\gamma$\ of the Bell
detection at the relay (see Ref.~\cite{CVMDIQKD}\ for more details). From this
shared post-relay CM, they can derive the secret-key rate of the protocol.

\section{General considerations on the security
analysis\label{GENERAL-CONSIDERATION}}

It is important to note that, once the shared CM\ $\mathbf{V}_{ab|\gamma}$ is
reconstructed by Alice and Bob, the secret-key rate can be (numerically)
computed no matter what the actual eavesdropping strategy was. In fact, it is
sufficient to consider the purification of the state $\rho_{ab|\gamma}$ into
an environment which is assumed to be fully controlled by Eve. This is a
pretty standard method in CV-QKD.

However, while this approach is valid for experimental demonstrations, it is
generally not sufficient for deriving analytic expressions of the key rate
$R$, just because there are too many free parameters in the CM. Having simple
analytic expressions is crucial in order to theoretically compare the
performances of different QKD protocols. The next theoretical step is
therefore the reduction of the free parameters to a minimum set which is
accessible to the parties and that allows us to write a closed formula for $R$
(or a lower-bound to $R$).

It is typical to derive a single quantifier of the noise, the so-called
`excess noise' $\varepsilon$, to be associated to the observed values of the
transmissivities $\tau_{A}$ and $\tau_{B}$. Such a reduction is the
non-trivial part of the theoretical analysis since it requires a minimization
of the rate with respect to all degrees of freedom of Eve, once that the
triplet $\tau_{A}$, $\tau_{B}$, and $\varepsilon$ has been fixed. One
important pre-requisite for such a reduction is the correct modelling of the
most general attack that Eve can perform against the protocol. The entire
\textquotedblleft space of the attacks\textquotedblright\ must be covered in
this analysis. As pointed out in Ref.~\cite{CVMDIQKD}, CV-MDI-QKD requires the
explicit consideration of all two-mode Gaussian attacks, not just one-mode
Gaussian attacks, where $g=g^{\prime}=0$. The latter class is in fact
restricted and can only lead to partial security proofs.

\section{Simple counter-example to one-mode attacks\label{COUNTER-EXAMPLE}}

Here we easily show that one-mode Gaussian attacks represent a restricted
class and, therefore, any security proof of CV-MDI-QKD based on these attacks
can only be partial. Furthermore, since they form a restricted class, it does
not make sense to claim their optimality.

For the sake of simplicity, consider the symmetric configuration~\cite{Carlo},
where Alice's and Bob's channels are identical lossy channels, with the same
transmissivity $\tau$. Extension to asymmetric configurations is just a matter
of technicalities. After the action of the relay, the shared CM of Alice and
Bob is simply given by~\cite{CVMDIQKD}%
\begin{equation}
\mathbf{V}_{ab|\gamma}=\left(
\begin{array}
[c]{cc}%
\mu\mathbf{I} & \mathbf{0}\\
\mathbf{0} & \mu\mathbf{I}%
\end{array}
\right)  -(\mu^{2}-1)\tau\left(
\begin{array}
[c]{cccc}%
\frac{1}{\theta} & 0 & -\frac{1}{\theta} & 0\\
0 & \frac{1}{\theta^{\prime}} & 0 & \frac{1}{\theta^{\prime}}\\
-\frac{1}{\theta} & 0 & \frac{1}{\theta} & 0\\
0 & \frac{1}{\theta^{\prime}} & 0 & \frac{1}{\theta^{\prime}}%
\end{array}
\right)  ,
\end{equation}
where
\begin{equation}
\theta:=2\left[  \tau\mu+(1-\tau)x\right]  ,~~\theta^{\prime}:=2\left[
\tau\mu+(1-\tau)x^{\prime}\right]  .
\end{equation}
with
\begin{equation}
x=\frac{\omega_{A}+\omega_{B}}{2}-g,~~x^{\prime}=\frac{\omega_{A}+\omega_{B}%
}{2}+g^{\prime}. \label{xPARA}%
\end{equation}

In the previous CM, the modulation parameter $\mu$ is known to Alice and Bob,
and also the transmissivity $\tau$ which is derived by comparing the shared
data and computing the first-order moments. By contrast, Alice and Bob do not
directly access the values of the thermal noise and the correlation
parameters, since they are combined in the $x$-parameters of Eq.~(\ref{xPARA}%
). The fact that the parameters $\omega_{A}$, $\omega_{B}$, $g$ and
$g^{\prime}$ get scrambled in $x$ and $x^{\prime}$ has led some
authors~\cite{authors} to claim that one-mode attacks ($g=g^{\prime}=0$) with
suitable values of the thermal noise ($\omega_{A}$ and $\omega_{B}$) could
simulate any two-mode attack with arbitrary $\omega_{A}$, $\omega_{B}$, $g$,
and $g^{\prime}$. However, it is quite trivial to check that this is not the case.

To understand this point, it is important to note that the components of the
CM $\mathbf{V}_{ab|\gamma}$ are monotonic in $x$ and $x^{\prime}$. As an
example, the top-left component%
\begin{equation}
\mathbf{V}_{11}=\mu-\frac{(\mu^{2}-1)\tau}{2\left[  \tau\mu+(1-\tau)x\right]
}%
\end{equation}
is increasing in $x$, so that $\mathbf{V}_{11}$\ is minimum when $x$ is
minimum. In the case of one-mode attacks ($g=g^{\prime}=0$), we have $x\geq1$.
It is therefore clear that any two-mode attack such that $x<1$ cannot be
simulated by one-mode attacks. Indeed there is an infinite number of such
two-mode attacks. In fact, let us assume that Eve performs a two-mode attack
with $\omega_{A}=\omega_{B}=\omega$ and $g^{\prime}=-g$. In this case, we have
$x=x^{\prime}=\omega-g$, and the condition $x<1$ corresponds to imposing
\begin{equation}
\omega-1<g\leq\sqrt{\omega^{2}-1}, \label{EntATT}%
\end{equation}
which are attacks where Eve's ancillas are entangled. \begin{figure}[ptbh]
\vspace{+0.15cm}
\par
\begin{center}
\includegraphics[width=0.375\textwidth] {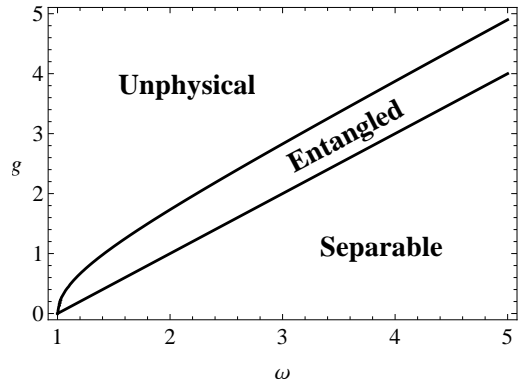}
\end{center}
\par
\vspace{-0.3cm}\caption{Correlation parameter $g$ versus thermal noise
$\omega$. For any value of $\omega$, two-mode attacks with $0\leq g\leq
\omega-1$ are performed with separable ancillas, while two-mode attacks
satisfying the stronger condition of Eq.~(\ref{EntATT}) have entangled
ancillas and cannot be simulated by one-mode attacks. Values of $g$ above
$\sqrt{\omega^{2}-1}$ are prohibited by quantum mechanics. }%
\label{figure}%
\end{figure}

Thus, for any value of $\omega$, we can pick an entangled two-mode attack
which cannot be simulated by one-mode attacks. In other words, this entangled
attack generates a shared CM $\mathbf{V}_{ab|\gamma}$ which does not belong to
the set of possible CMs associated with one-mode attacks. As depicted in
Fig.~\ref{figure}, there is an infinite number of entangled attacks which
cannot be reduced to one-mode attacks.

One may attempt to enlarge the set of one-mode attacks by allowing for
squeezed thermal noise, i.e., the use of thermal states with asymmetric
variances, $\omega_{A}^{q}$ for the $q$-quadrature and $\omega_{A}^{p}$ for
the $p$-quadrature (and similarly, $\omega_{B}^{q}$ and $\omega_{B}^{p}$, for
the other ancilla). In this case, Eq.~(\ref{xPARA}) for one-mode attacks would
become%
\begin{equation}
x=\frac{\omega_{A}^{q}+\omega_{B}^{q}}{2},~~x^{\prime}=\frac{\omega_{A}%
^{p}+\omega_{B}^{p}}{2}~.
\end{equation}
\newline However, since $\omega_{A}^{q}\omega_{A}^{p}\geq1$ and $\omega
_{B}^{q}\omega_{B}^{p}\geq1$, it is easy to check that realizing $x<1$ would
imply $x^{\prime}>1$, and vice versa. As a result, there will always be
components in the shared CM $\mathbf{V}_{ab|\gamma}$ whose values, for
entangled attacks, cannot be realized by assuming one-mode attacks.

\section{Conclusion\label{CONCLUSIONS}}

We have considered the security analysis of CV-MDI-QKD. We have explicitly
shown that one-mode Gaussian (entangling-cloner) attacks represent a
restricted class, which cannot generate all the possible shared CMs for Alice
and Bob. This is true for any fixed value of the transmissivity $\tau$ for the
two lossy channels (extension to different transmissivities $\tau_{A}$ and
$\tau_{B}$ is trivial). This very simple result confirms the necessity of
explicitly studying two-mode Gaussian attacks in the security analysis of
CV-MDI-QKD, as originally considered in Ref.~\cite{CVMDIQKD}.
%At the same time, it exposes a
%fatal flaw in~? and the incompleteness of security proofs such as that in
%Ref.~\cite{MaXC}.

\textbf{Acknowledgments}.~This work has been supported by the EPSRC `UK
Quantum Communications HUB' (EP/M013472/1).


\begin{thebibliography}{99}
\bibitem {EBrepr}In an entanglement-based representation, Alice has an EPR
state~\cite{RMP} $\rho_{aA}$ of modes $a$\ and $A$. By heterodyning mode $a$,
she probabilistically prepares coherent states on mode $A$. Similarly, Bob has
another EPR state $\rho_{bB}$.
\end{thebibliography}
\end{document}